\documentclass[12pt,letterpaper]{article}

\title{Equal charge black holes and seven dimensional gauged supergravity}
\author{
David D. K. Chow
}
\date{}

\usepackage{amsmath}
\usepackage{amsfonts}
\usepackage{amssymb}
\usepackage{amscd}
\usepackage{latexsym}

\makeatletter
\@addtoreset{equation}{section}
\makeatother

\newcommand{\ben}{\begin{equation}} \newcommand{\een}{\end{equation}}
\newcommand{\bea}{\setlength\arraycolsep{2pt} \begin{eqnarray}}
\newcommand{\eea}{\end{eqnarray}} \newcommand{\nnr}{\nonumber \\}
\newcommand{\pd}{\partial} \newcommand{\ud}{\textrm{d}}
\newcommand{\ui}{\textrm{i}}
\newcommand{\ue}{\textrm{e}}

\begin{document}
\begin{titlepage}
\begin{flushright}
DAMTP-2007-110
\end{flushright}
\vspace*{100pt}
\begin{center}
{\bf \Large{Equal charge black holes and seven dimensional gauged supergravity}}\\
\vspace{50pt}
\large{David D. K. Chow}
\end{center}
\begin{center}
{\it Department of Applied Mathematics and Theoretical Physics, University of Cambridge,\\
Centre for Mathematical Sciences, Wilberforce Road, Cambridge CB3 0WA, UK}\\
{\tt D.D.K.Chow@damtp.cam.ac.uk}\\
\vspace{50pt}
{\bf Abstract\\}
\end{center}
We present various supergravity black holes of different dimensions with some $\textrm{U}(1)$ charges set equal in a simple, common form.  Black hole solutions of seven
dimensional $\textrm{U}(1)^2$ gauged supergravity with three
independent angular momenta and two equal $\textrm{U}(1)$ charges are
obtained.  We investigate the thermodynamics and the BPS limit of this
solution, and find that there are rotating supersymmetric black holes without naked closed timelike
curves.  There are also supersymmetric topological soliton solutions
without naked closed timelike curves that have a smooth geometry.
\end{titlepage}

\section{Introduction}

There has recently been interest in solutions of gauged supergravity theories because of the AdS/CFT correspondence \cite{AdSCFT, gaugecorncst, AdShol, AdSCFTPhysRep}.  Using exact solutions from the gravitational side of the correspondence, one hopes to learn about the dual gauge theory.  Supersymmetric black hole solutions are useful for
comparing the thermodynamics on both sides of the correspondence,
since non-renormalization theorems mean that some results on the
conformal field theory side may be extrapolated from weak to strong
coupling.  More general non-extremal black hole solutions have a
non-zero Hawking temperature and are useful for studying the dual
field theory at non-zero temperature.

It is of interest to consider not only four spacetime dimensions, but also higher spacetime dimensions $D > 4$, since the AdS/CFT correspondence may be studied in a variety of dimensions.  Of particular interest are
supergravity theories in spacetime dimensions $D = 4, 5, 7$, since the AdS/CFT
correspondence relates these to superconformal field theories on a
large number, $N$, of M2-branes, D3-branes, and M5-branes
respectively; these branes preserve 32 supercharges and have maximal supersymmetry.  More precisely, there is a duality between M-theory on
$\textrm{AdS}_4 \times S^7$ and a non-abelian $D = 3$, $\mathcal{N} = 8$
superconformal theory, between type IIB string theory on
$\textrm{AdS}_5 \times S^5$ and $D = 4$, $\mathcal{N} = 4$,
$\textrm{SU}(N)$ super Yang--Mills theory, and between M-theory on
$\textrm{AdS}_7 \times S^4$ and a non-abelian $D= 6$, $\mathcal{N} =
(2, 0)$ superconformal theory.  These respectively correspond to the
maximal $D = 4$, $\mathcal{N} =8$, $\textrm{SO}(8)$; $D = 5$, $\mathcal{N} = 8$, $\textrm{SO}(6)$; and $D =
7$, $\mathcal{N} = 4$, $\textrm{SO}(5)$ gauged supergravities, which have respective Cartan
subgroups $\textrm{U}(1)^4$, $\textrm{U}(1)^3$ and
$\textrm{U}(1)^2$.

There has been much progress over the last few years in obtaining new,
non-extremal, asymptotically AdS black hole solutions of gauged supergravity theories in four \cite{crotbh4d},
five \cite{crotbh5dU13, 5dgsugrabhind, bh5dgsugra,
  nonextremrotbh5dgsugra, newrotbh5d} and seven \cite{crotbh7d} dimensions.  However, in each case, we
have not yet obtained a non-extremal solution with all rotation and
charge parameters arbitrary.  In addition to the non-extremal black hole families of solutions in the literature, there are also some
known supersymmetric solutions, which should belong to a larger family
of non-extremal solutions yet to be discovered.  For the five
dimensional case, building on previous work on supersymmetric $\textrm{AdS}_5$
black holes \cite{susyAdS5bh, gensusyAdS5bh}, a supersymmetric solution with two rotation and
three charge parameters arbitrary except for a single BPS constraint
is known \cite{susymcAdS5}.  The basis for those studies is the
classification of supersymmetric solutions using the $G$-structure
formalism, introduced in the context of five
dimensional minimal gauged supergravity by \cite{5dgaugedsugra}.
There has been some work which extends the supersymmetric
classification to higher dimensions, for example in the dimension we
focus on in this paper, $D = 7$ \cite{tKillspin7d}, however the classification becomes increasingly
implicit.  Nevertheless, supersymmetric solutions have been successfully found as a limit of non-extremal solutions without such a formalism, for example amongst the non-extremal solutions cited above.

In this paper, we consider certain supergravity black hole solutions
that are both charged and rotating, but which have the simplification
that we truncate to the Cartan subgroup, which is of the form $\textrm{U}(1)^k$, of the full gauge group, and some combinations of these
$\textrm{U}(1)$ charges are set equal.  The main result of this paper is a new black hole solution of seven
dimensional gauged supergravity with three independent angular momenta
and two equal $\textrm{U}(1)$ charges.  However, for non-extremal
solutions of gauged supergravity theories, there is no known solution
generating technique, and instead one must rely on inspired
guesswork.  Therefore the ``derivation'' of the new solution will
consist of a presentation of some previously known solutions in a
unified manner, from which one might obtain a tight ansatz to find the
result.  In particular, in section 2, we review the relevant solutions of toroidally compactified
heterotic supergravity in \cite{nearBPSsat}, of four dimensional gauged
supergravity in \cite{crotbh4d}, and of five dimensional gauged supergravity in
\cite{5dgsugrabhind}.  The case of toroidally compactified heterotic supergravity is useful as its Lagrangian gives rise to the ungauged limit of the supergravity Lagrangians we shall consider, and global symmetries of the theory give a mechanical solution generating technique that charges up a neutral solution, and so provide the ungauged limits of the solutions we seek.  In section 3, we concentrate on the case of
seven dimensional gauged supergravity, in particular reviewing the solution of
\cite{crotbh7d}, which has equal rotation parameters, specializing to the case of two equal $\textrm{U}(1)$ charges in the $\textrm{U}(1)^2$ maximal abelian subgroup of the full
$\textrm{SO}(5)$ gauge group.  Guided by these previous solutions with equal charges, we then
write down a new solution of seven dimensional gauged supergravity
with three independent angular momenta and both $\textrm{U}(1)$
charges equal.  There is a brief discussion of the curvature singularities, which we find are similar to those of other higher dimensional black holes.  We examine the thermodynamics of the
solution and also the BPS limit, finding that the solution includes
supersymmetric black holes without naked closed timelike curves.  Like
in four and five dimensions, these supersymmetric black holes must
rotate.  We investigate the fraction of supersymmetry preserved by the supersymmetric black holes and find that they are all $\frac{1}{8}$ supersymmetric.  Of the more general supersymmetric solutions, naked closed
timelike curves may also be avoided by a class of topological soliton
solutions, which have a smooth geometry provided that the parameters
obey a certain quantization condition.  The topological soliton solutions are
generally $\frac{1}{8}$ supersymmetric, but we investigate whether the
supersymmetry can be enhanced, and find that it can in special cases.

\section{Equal charge black holes}

As a first step towards obtaining new seven dimensional gauged
supergravity black holes with equal charges, we first reexamine some
other supergravity black hole solutions with equal charges.  We first consider
solutions of ungauged supergravity, in particular those of toroidally
compactified heterotic supergravity, before reviewing the most
relevant solutions of four and five dimensional gauged supergravity.

\subsection{Toroidally compactified heterotic supergravity}

We first consider toroidally compactified heterotic supergravity,
since its Lagrangian gives rise to the ungauged limit of the gauged
supergravity theories we shall later consider.  Solutions of this
ungauged theory can be easier to obtain, and we shall later see
examples of how certain solutions generalize to the gauged theories.

The bosonic fields of heterotic supergravity are the graviton
$g_{ab}$, a dilaton $\phi$, a Yang--Mills field $A_a$ in the adjoint representation of $\textrm{E}_8 \times \textrm{E}_8$ or
$\textrm{Spin}(32) / \mathbb{Z}_2$, and a Kalb--Ramond two-form potential $B_{ab}$.  We
truncate to the $\textrm{U}(1)^{16}$ Cartan subgroup of the
Yang--Mills sector, which is a consistent bosonic truncation.  In Einstein frame, the ten dimensional bosonic Lagrangian of
heterotic supergravity is
\ben
\mathcal{L}_{10} = R \star 1 - \frac{1}{2} \star \ud \phi \wedge \ud \phi -
\frac{1}{2} \sum_{I=1}^{16} \ue ^{- \phi / 2} \star F_{(2)}^I \wedge
F_{(2)}^I - \frac{1}{2} \ue ^{- \phi} \star H_{(3)} \wedge H_{(3)} ,
\een
with $F_{(2)}^I = \ud A_{(1)}^I$ and $H_{(3)} = \ud B_{(2)} - \frac{1}{2} \sum_{I=1}^{16} F_{(2)}^I \wedge A_{(1)}^I$.  Only terms with up to two derivatives appear in this Lagrangian.

If we perform a Kaluza--Klein reduction on a torus $T^d$ to $D = 10 - d \geq 4$ dimensions, then we
obtain the $D$-dimensional low-energy effective Lagrangian, which is
of the form $\mathcal{L}_D = \mathcal{L}_{D,1} + \mathcal{L}_{D,2} +
\mathcal{L}_{D,3}$,
where
\bea
\mathcal{L}_{D,1} & = & R \star 1 - \frac{1}{2} \Bigg( \star \ud \phi
  \wedge \ud \phi + \sum_{i=1}^d \star \ud \phi_i \wedge \ud \phi_i + \sum_{i=1}^d \ue ^{{\bf C}_i \cdot
  {\boldsymbol \phi}} \star F_{i (2)} \wedge F_{i (2)} \nnr
&& \qquad \qquad \quad + \sum_{1 \leq i
< j \leq d} \ue ^{{\bf C}_{ij} \cdot {\boldsymbol \phi}} \star F_{ij (1)}
\wedge F_{ij (1)} \Bigg) , \nnr
\mathcal{L}_{D,2} & = & - \frac{1}{2}\sum_{I = 1}^{16} \left( \ue ^{{\bf A} \cdot
  {\boldsymbol \phi}} \star F^I_{(2)} \wedge F^I_{(2)} + \sum_{i=1}^d \ue ^{{\bf A}_i
  \cdot {\boldsymbol \phi}} \star F_{i (1)}^I \wedge F_{i (1)}^I
\right) , \nnr
\mathcal{L}_{D,3} & = & - \frac{1}{2} \Bigg( \ue ^{{\bf B} \cdot {\boldsymbol \phi}} \star
  H_{(3)} \wedge H_{(3)} + \sum_{i=1}^d \ue ^{{\bf B}_i \cdot {\boldsymbol \phi}}
  \star H_{i (2)} \wedge H_{i (2)} \nnr
&& \qquad + \sum_{1 \leq i < j \leq d} \ue ^{{\bf B}_{ij}
    \cdot {\boldsymbol \phi}} \star H_{ij (1)} \wedge H_{ij (1)}
\Bigg) .
\eea
$\mathcal{L}_{D,1}$ contains the ten dimensional
scalar and terms that arise from reduction of the Einstein--Hilbert
term, including $d$ dilatons, $d$ vectors and, from reduction of the
vectors, $\frac{1}{2} d(d-1)$ so-called ``axions'', which are scalars that arise
from off-diagonal metric components.  $\mathcal{L}_{D,2}$ comes from
reduction of the two-form field strengths and $\mathcal{L}_{D,3}$ comes from reduction of
the three-form field strength.  ${\boldsymbol \phi} = (\phi, \phi_1 ,
\ldots \phi_d)$ is a vector field with its $d+1$ components being the
ten dimensional scalar and the $d$ dilatons.  ${\bf A}$, ${\bf A}_i$,
${\bf B}$, ${\bf B}_i$, ${\bf B}_{ij}$,
${\bf C}_i$ and ${\bf C}_{ij}$ are constant vectors with $d+1$
components, which are related to the root lattice of $\textrm{O}(10 - D, 26 - D)$.  We shall not require the full expressions for the
constant vectors or for the field strengths
in terms of potentials, which may be found in \cite{MhetdualKK} (see also \cite{dualdual} for the procedure applied to eleven dimensional supergravity), except for the relation ${\bf B} = 2 {\bf A}$.

We shall consider the consistent bosonic truncation to the sector in which the only fields
turned on are one linear combination of the dilatons,
$\varphi = {\bf A} \cdot {\boldsymbol \phi} \sqrt{(D-2) / 2}$, two equal
one-form potentials, $A_{(1)} = A_{(1)}^1 = A_{(1)}^2$, and the
two-form potential $B_{(2)}$.  The resulting field strengths are
$F_{(2)} = \ud A_{(1)}$ and $H_{(3)} = \ud B_{(2)} - A_{(1)} \wedge
\ud A_{(1)}$, in terms of which a Lagrangian for the field equations
is
\ben
\mathcal{L}_D = R \star 1 - \frac{1}{2} \star \ud \varphi \wedge \ud
\varphi - \ue ^{2 \varphi / \sqrt{2(D-2)}} \star F_{(2)} \wedge F_{(2)} -
\frac{1}{2} \ue ^{4 \varphi / \sqrt{2(D-2)}} \star H_{(3)} \wedge H_{(3)} .
\label{hetL}
\een

To make contact with other supergravity theories, we obtain an
equivalent Lagrangian by dualizing the three-form field strength, $H_{(3)}$, in favour
of a $(D-3)$-form field strength, $F_{(D-3)}$.  The Poincar\'{e}
dualization procedure first involves noting the Bianchi identity for
the three-form field strength, $\ud H_{(3)} + F_{(2)} \wedge F_{(2)} = 0$,
which is imposed by adding to the Lagrangian the term $(-1)^{D-1}
A_{(D-4)} \wedge (\ud H_{(3)} + F_{(2)} \wedge F_{(2)})$, treating
$A_{(D-4)}$ as a Lagrange multiplier.  Varying the modified Lagrangian
with respect to $H_{(3)}$, the algebraic equation of motion for
$H_{(3)}$ gives an expression for the dual field strength,
\ben
F_{(D-3)} = \ud A_{(D-4)} = \ue ^{4 \varphi / \sqrt{2(D-2)}} \star
H_{(3)} .
\een
Substituting back into the original Lagrangian and integrating by
parts, we obtain the dual Lagrangian
\bea
\mathcal{L}_D & = & R \star 1 - \frac{1}{2} \star \ud \varphi \wedge \ud
\varphi - \ue ^{2 \varphi / \sqrt{2(D-2)}} \star F_{(2)} \wedge F_{(2)} -
\frac{1}{2} \ue ^{-4 \varphi / \sqrt{2(D-2)}} \star F_{(D-3)} \wedge
F_{(D-3)} \nnr
&& + (-1)^{D-1} F_{(2)} \wedge F_{(2)} \wedge A_{(D-4)} ,
\label{dualhetL}
\eea
where $F_{(2)} = \ud A_{(1)}$ and $F_{(D-3)} = \ud A_{(D-4)}$.  The
Chern--Simons terms appearing originate from the dependence of
$H_{(3)}$ on $A_{(1)}$.  We have chosen not to rescale $F_{(2)}$ to the canonical normalization
for a single field, again for convenience when comparing with other
supergravity theories.

\subsection{Equal charge black holes of toroidally compactified heterotic supergravity}

From uncharged black holes of Einstein gravity in the absence of a
cosmological constant, one may generate charged generalizations in the
context of toroidally compactified heterotic supergravity as a result
of global symmetries of the theory.  For rotating solutions in higher dimensions,
the procedure was used in \cite{nearBPSsat} to obtain charged and
rotating black hole solutions from the neutral Myers--Perry solution \cite{bhhigherdim}.
We shall see that the charged solution simplifies substantially in the
case that the charge parameters are equal, which will form the basis
of generalizations to equal charge solutions of gauged supergravity
theories.

The most illuminating way of writing the solution in the equal charge
case is to use latitudinal and azimuthal coordinates that generalize
the coordinates used for the Pleba\'{n}ski solution \cite{classsolEMeq} (see also \cite{PlebDem} for the inclusion, in four dimensions, of an acceleration parameter, as in the C-metric), through which the solution takes a
rather symmetrical form, and to write the metric using a set of simple
vielbeins.  In higher dimensions, such an approach was used in
\cite{genKNUTAdSalld} to obtain NUT charge generalizations of the
higher dimensional Kerr--AdS solution.  There, the key observation was
that for the round metric on a unit sphere $S^{D-2}$,
\ben
\ud \Omega_{D-2}^2 = \sum_{i=1}^{\lfloor D/2 \rfloor} \ud \mu_i^2 +
\sum_{i=1}^{\lfloor (D-1)/2 \rfloor} \mu_i^2 \ud \phi_i^2 , \quad
\sum_{i=1}^{\lfloor D/2 \rfloor} \mu_i^2 = 1 ,
\een
the latitudinal coordinates $\mu_i$ may be parameterized as
\ben
\mu_i^2 = \frac{\prod_{\alpha = 1}^{n-1} (a_i^2 -
  y_\alpha^2)}{\prod^\prime{_{k=1}^n} (a_i^2 - a_k^2)} ,
\label{mui}
\een
where $D = 2n$ for even dimensions and $D = 2n+1$ for odd dimensions.
We have used the notation $\prod '$ to indicate that we exclude the factor that vanishes from a product.  The round metric then takes a diagonal form with
\ben
\sum_{i=1}^n \ud \mu_i^2 = (-1)^{n+1} \sum_{\alpha = 1}^{n-1} \frac{y_\alpha^2 \prod^\prime{_{\beta = 1}^{n-1}} (y_\alpha^2 -
  y_\beta^2)}{\prod_{k=1}^n (a_k^2 - y_\alpha^2)} \ud y_\alpha^2 .
\een
For the azimuthal coordinates, the higher dimensional generalization
of Boyer--Lindquist coordinates $\phi_i$ retain the property that they
are periodic with canonical normalization so that their period is $2
\pi$.  For computational purposes and conciseness, it is more
convenient to perform a linear coordinate transformation of the
azimuthal coordinates $\phi_i$ and the Boyer--Lindquist time coordinate
$t$ to a higher dimensional generalization of those
used by Pleba\'{n}ski, as we shall later use and denote by $\psi_i$; the coordinate change
may be found in \cite{genKNUTAdSalld}.  Readers unfamiliar with this
coordinate system for higher dimensional black holes may find it
helpful to look at more specific examples first, before considering
arbitrary dimensions, for example in \cite{genKNUTAdSalld}, where
explicit expressions for the Kerr--NUT--AdS solution in six and seven
dimensions may be found, and the solutions of four and five
dimensional gauged supergravity that we review later.

The latitudinal coordinates $y_\alpha$ were first introduced
by Jacobi \cite{Jacobi}, so I suggest calling them Jacobi coordinates (although the $n = 3$ case was previously considered by Neumann in analysing the three dimensional harmonic oscillator constrained on $S^2$ \cite{Neumann}).  Because of
the use of these azimuthal coordinates by Carter for expressing the
Kerr--AdS solution \cite{Carter}, I suggest that the $\psi_i$ coordinates
be called Carter coordinates.  Although Pleba\'{n}ski
\cite{classsolEMeq} suggested the name of
Boyer coordinates for the full set of all four coordinates, such
terminology has not caught on, perhaps because of possible confusion
with Boyer--Lindquist coordinates.  I therefore instead suggest that the full set of coordinates $(y_\alpha , \psi_i)$
be called Jacobi--Carter coordinates.

We should note that there are some typographical errors in the general
solution of \cite{nearBPSsat}, noted for example in \cite{crotbh7d}.
Also, compared to \cite{nearBPSsat}, we have changed the sign of
$\phi_i$ and set $l_i = a_i$.

\subsubsection{Even dimensions $D = 2n$}

In Boyer--Lindquist coordinates, the solution of \cite{nearBPSsat} with both $\textrm{U} (1)$ charges set equal in even dimensions $D = 2n$ may be written as
\bea
&& \ud s^2 = H^{2/(D-2)} \Bigg\{ - \frac{R}{H^2 U}
\mathcal{A}^2 + \frac{U}{R} \ud r^2 \nnr
&& \qquad \qquad \qquad \quad + \sum_{\alpha = 1}^{n-1}  \Bigg[ \frac{X_\alpha}{U_\alpha} \left( \ud t - \sum_{i=1}^{n-1} \frac{(r^2 +
  a_i^2) \gamma_i}{a_i^2 - y_\alpha^2} \ud \tilde{\phi}_i - \frac{2 m
  s^2 r}{HU} \mathcal{A} \right) ^2 + \frac{U_\alpha}{X_\alpha} \ud
y_\alpha^2 \Bigg] \Bigg\} , \nnr
&& \ue ^{\varphi '} = \frac{1}{H} , \quad A_{(1)} = \frac{2mscr}{HU}
\mathcal{A}, \quad B_{(2)} = \frac{2 m s^2 r}{HU} \ud t \wedge
\sum_{i=1}^{n-1} \gamma_i \ud \tilde{\phi}_i ,
\label{evengeneralBL}
\eea
where
\bea
&& U = \prod_{\alpha = 1}^{n-1} (r^2 + y_\alpha^2) , \quad U_\alpha = -
(r^2 + y_\alpha^2) \sideset{}{'}\prod_{\beta = 1}^{n-1} (y_\beta^2 -
y_\alpha^2) , \quad \gamma_i = \prod_{\alpha = 1}^{n-1} (a_i^2 -
y_\alpha^2) , \nnr
&& R = \prod_{k=1}^{n-1} (r^2 + a_k^2) - 2mr , \quad
X_\alpha = - \prod_{k=1}^{n-1} (a_k^2 - y_\alpha^2) , \quad
\mathcal{A} = \ud t - \sum_{i=1}^{n-1} \gamma_i \ud \tilde{\phi}_i ,
\nnr
&& H = 1 + \frac{2 m
  s^2 r}{U} , \quad s = \sinh \delta , \quad c = \cosh \delta , \quad \tilde{\phi}_i = \frac{\phi_i}{a_i
  \prod^\prime{_{k=1}^{n-1}} (a_i^2 - a_k^2)} ,
\eea
and the normalization of the scalar of \cite{nearBPSsat}, $\varphi '$, is related to the
scalar that we have been using, $\varphi$, by $\varphi ' = \sqrt{(D-2)/2} \varphi$.

The metric takes a slightly simpler form if we make a linear
coordinate transformation of the azimuthal coordinates $\phi_i$ to
Carter coordinates $\psi_i$.  In these coordinates, there is a compact expression for the solution if we make analytic continuations to give a Riemannian metric.  We analytically continue the radial coordinate $r$, so that it
appears on an equal footing as the other coordinates $y_\alpha$, and define $n$ coordinates $x_\mu$ by
\bea
x_\alpha & = & y_\alpha , \quad 1 \leq \alpha \leq n-1 , \nnr
x_n & = & \ui r .
\label{xmudef}
\eea
To keep the metric real, we also make the analytic continuation
$m_n = - \ui m$.  The new coordinate $t' = \psi_0$, which would be a time coordinate in Lorentzian signature, may be placed on a similar footing
as the latitudinal coordinates $\psi_i$.  It is also convenient to
record the dual potential $A_{(D-4)}$ rather than $B_{(2)}$.  The
solution takes the form
\bea
&& \ud s^2 = H^{2/(D-2)} \sum_{\mu = 1}^n \left[ \frac{X_\mu}{U_\mu}
  \left( \mathcal{A}_\mu - \frac{2 m_n s^2 x_n}{H U_n} \mathcal{A}_n
  \right) ^2 + \frac{U_\mu}{X_\mu} \ud x_\mu^2 \right] , \nnr
&& \ue ^{\varphi '} = \frac{1}{H} , \quad A_{(1)} = \frac{2 m_n s c x_n}{H
  U_n} \mathcal{A}_n , \nnr
&& A_{(D-4)} = \frac{2 \ui m_n s^2 \prod_{\alpha = 1}^{n-1}
  x_\alpha}{(n-2)! U_n} \left(
\sum_{\alpha = 1}^{n-1} \frac{x_\alpha^2 - x_n^2}{x_\alpha} \ud
x_\alpha \wedge \mathcal{A}_{\alpha n} \right) ^{n-2} ,
\label{evengeneralPleb}
\eea
where
\bea
&& U_\mu = \sideset{}{'} \prod_{\nu = 1}^n (x_\nu^2 - x_\mu^2) , \quad
X_\mu = - \prod_{k=1}^{n-1} (a_k^2 - x_\mu^2) + 2 m_\mu x_\mu ,  \quad
m_\mu = m_n \delta_{\mu n} , \nnr
&& \mathcal{A}_\mu = \sum_{k=0}^{n-1} A_\mu^{(k)} \ud \psi_k , \quad A_\mu^{(k)} = \sum_{\substack{\nu_1 < \nu_2 < \ldots < \nu_k \\
    \nu_i \neq \mu}} x_{\nu_1}^2 x_{\nu_2}^2 \ldots x_{\nu_k}^2 , \nnr
&& \mathcal{A}_{\mu \nu} = \sum_{k=1}^{n-1} A_{\mu
  \nu}^{(k-1)} \ud \psi_k , \quad A_{\mu \nu}^{(k)} = \sum_{\substack{\nu_1 < \nu_2 < \ldots < \nu_k
    \\ \nu_i \neq \mu , \nu}} x_{\nu_1}^2 x_{\nu_2}^2 \ldots
x_{\nu_k}^2 , \nnr
&& H = 1 + \frac{2 m_n s^2 x_n}{U_n} , \quad s = \sinh \delta, \quad c = \cosh \delta .
\eea

\subsubsection{Odd dimensions $D = 2n+1$}

In Boyer--Lindquist coordinates, the solution of \cite{nearBPSsat} with both $\textrm{U} (1)$ charges set equal in odd dimensions $D = 2n+1$ may be written as
\bea
&& \ud s^2 = H^{2/(D-2)} \Bigg\{ - \frac{R}{H^2 U}
\mathcal{A} ^2 + \frac{U}{R} \ud r^2 \nnr
&& \qquad \qquad \qquad \quad + \sum_{\alpha = 1}^{n-1}
\Bigg[ \frac{X_\alpha}{U_\alpha} \left( \ud t - \sum_{i=1}^n \frac{(r^2
    + a_i^2) \gamma_i}{a_i^2 - y_\alpha^2} \ud \tilde{\phi}_i -
  \frac{2 m s^2}{HU} \mathcal{A} \right) ^2 + \frac{U_\alpha}{X_\alpha} \ud y_\alpha^2 \Bigg] \nnr
&& \qquad \qquad \qquad \quad + \frac{\prod_{k=1}^n a_k^2}{r^2
  \prod_{\alpha = 1}^{n-1} y_\alpha^2} \left( \ud t - \sum_{i=1}^n
  \frac{(r^2 + a_i^2) \gamma_i}{a_i^2} \ud \tilde{\phi}_i - \frac{2 m s^2}{HU} \mathcal{A} \right) ^2 \Bigg\} , \nnr
&& \ue ^{\varphi '} = \frac{1}{H} , \quad A_{(1)} = \frac{2msc}{HU}
\mathcal{A} , \quad B_{(2)} = \frac{2 m s^2}{HU} \ud t \wedge
\sum_{i=1}^n \gamma_i \ud \tilde{\phi}_i ,
\label{oddgeneralBL}
\eea
where
\bea
&& U = \prod_{\alpha = 1}^{n-1} (r^2 + y_\alpha^2) , \quad U_\alpha =
- (r^2 + y_\alpha^2) \sideset{}{'} \prod_{\beta = 1}^{n-1} (y_\beta^2
- y_\alpha^2) , \quad \gamma_i = a_i^2 \prod_{\alpha = 1}^{n-1} (a_i^2 -
y_\alpha^2) , \nnr
&& R = \frac{1}{r^2} \prod_{k=1}^n (r^2 + a_k^2) - 2m , \quad X_\alpha
= \frac{1}{y_\alpha^2} \prod_{k=1}^n (a_k^2 - y_\alpha^2) , \quad
\mathcal{A} = \ud t - \sum_{i=1}^n \gamma_i \ud \tilde{\phi}_i , \nnr
&& H = 1 + \frac{2 m
  s^2}{U} , \quad s = \sinh \delta , \quad c = \cosh \delta , \quad \tilde{\phi}_i = \frac{\phi_i}{a_i
  \prod^\prime{_{k=1}^n} (a_i^2 - a_k^2)} ,
\eea
again with $\varphi ' = \sqrt{(D-2)/2} \varphi$.

We again analytically continue the radial coordinate $r$ for
convenience when using Carter coordinates, using the same
definition for the $n$ coordinates $x_\mu$ as for the even dimensional
case in (\ref{xmudef}).  The solution after these analytic continuations is
\bea
&& \ud s^2 = H^{2/(D-2)} \Bigg\{ \sum_{\mu = 1}^n \left[ \frac{X_\mu}{U_\mu}
  \left( \mathcal{A}_\mu - \frac{2 m_n s^2}{H
      U_n} \mathcal{A}_n \right) ^2 + \frac{U_\mu}{X_\mu} \ud x_\mu^2
\right] \nnr
&& \qquad \qquad \qquad \quad - \frac{\prod_{i=1}^n a_i^2}{\prod_{\mu = 1}^n x_\mu^2} \left(
 \sum_{k=0}^n A^{(k)} \ud \psi_k - \frac{2 m_n s^2}{H U_n}
 \mathcal{A}_n \right) ^2 \Bigg\} , \nnr
&& \ue ^{\varphi '} = \frac{1}{H} , \quad A_{(1)} = \frac{2 m_n s c}{H U_n}
\mathcal{A}_n , \nnr
&& A_{(D-4)} = \frac{2 m_n s^2 \prod_{i=1}^n a_i}{(n-2)! U_n} \sum_{k=1}^n A_n^{(k-1)} \ud \psi_k \wedge
\left( \sum_{\alpha = 1}^{n-1} \frac{x_\alpha^2 - x_n^2}{x_\alpha} \ud x_\alpha \wedge \mathcal{A}_{\alpha
    n} \right) ^{n-2} ,
\label{oddgeneralPleb}
\eea
where
\bea
&& U_\mu = \sideset{}{'} \prod_{\nu = 1}^n (x_\nu^2 - x_\mu^2) , \quad
X_\mu = \frac{1}{x_\mu^2} \prod_{k=1}^n (a_k^2 - x_\mu^2) + 2 m_\mu ,
\quad m_\mu = m_n \delta_{\mu n} , \nnr
&&  A^{(k)} = \sum_{\nu_1 < \nu_2 < \ldots < \nu_k} x_{\nu_1}^2 x_{\nu_2}^2 \ldots x_{\nu_k}^2 , \\
&& \mathcal{A}_\mu = \sum_{k=0}^{n-1} A_\mu^{(k)} \ud \psi_k , \quad A_\mu^{(k)} = \sum_{\substack{\nu_1 < \nu_2 < \ldots < \nu_k \\
    \nu_i \neq \mu}} x_{\nu_1}^2 x_{\nu_2}^2 \ldots x_{\nu_k}^2 , \nnr
&& \mathcal{A}_{\mu \nu} = \sum_{k=1}^{n-1} A_{\mu \nu}^{(k-1)}
\ud \psi_k , \quad A_{\mu \nu}^{(k)} = \sum_{\substack{\nu_1 < \nu_2 < \ldots < \nu_k \\
    \nu_i \neq \mu , \nu}} x_{\nu_1}^2 x_{\nu_2}^2 \ldots x_{\nu_k}^2
, \nnr
&& H = 1 + \frac{2
  m_n s^2}{U_n} , \quad s = \sinh \delta , \quad c = \cosh \delta .
\eea

\subsection{Four dimensional gauged supergravity}

The maximal $D = 4$, $\mathcal{N} = 8$, $\textrm{SO}(8)$ gauged
supergravity may be obtained by dimensional reduction of eleven
dimensional supergravity on $S^7$.  Truncating so that we only include
gauge fields in the $\textrm{U}(1)^4$ Cartan subgroup of the full
gauge group, we arrive at $\mathcal{N} = 2$ gauged supergravity
coupled to three vector multiplets.  The bosonic fields are the
graviton, four $\textrm{U}(1)$ gauge fields, three dilatons and three
axions.

Considering the truncation to pairwise equal charges, the bosonic Lagrangian is
\bea
\mathcal{L}_4 & = & R \star 1 - \frac{1}{2} \star \ud \varphi \wedge \ud
\varphi - \frac{1}{2} \ue ^{- 2 \varphi} \star \ud \chi \wedge \ud \chi -
\frac{1}{2} \ue ^{\varphi} (\star F_{(2)}^1 \wedge F_{(2)}^1 + \star
F_{(2)}^2 \wedge F_{(2)}^2) \nnr
&& - \frac{1}{2} \chi (F_{(2)}^1 \wedge F_{(2)}^1 + F_{(2)}^2 \wedge
F_{(2)}^2) + g^2 (4 + 2 \cosh \varphi + \ue ^{- \varphi} \chi^2) \star 1
.
\eea
Compared with \cite{crotbh4d}, we have changed the sign of $\varphi$,
and adjusted the sign of the potential so that setting both scalars to
zero gives a negative cosmological constant.

Black hole solutions of this truncation of the gauged supergravity
theory were obtained in \cite{crotbh4d}.  They are parameterized by
the angular momentum, two independent $\textrm{U}(1)$ charges, mass
and NUT charge, although we do not consider NUT charge here.  In the
ungauged limit, the solutions can be obtained from a more general known four
charge solution by setting the charges to be pairwise equal.

A further consistent bosonic truncation is to take $A_{(1)}^1 = 0$.
The field equations can be obtained from the Lagrangian
\bea
\mathcal{L}_4 & = & R \star 1 - \frac{1}{2} \star \ud \varphi \wedge \ud
\varphi - X^{-2} \star F_{(2)} \wedge F_{(2)} - \frac{1}{2} X^4 \star
\ud \chi \wedge \ud \chi - \chi F_{(2)} \wedge F_{(2)} \nnr
&& + g^2 (X^2 + 4 + X^{-2} + X^2 \chi^2) \star 1 ,
\eea
where $X = \ue ^{- \varphi / 2}$ and $F_{(2)} = \ud A_{(1)} = (1 /
\sqrt{2}) \ud A_{(1)}^2$.  Setting $g = 0$, we recover the Lagrangian
of (\ref{dualhetL}).

The black hole solution without NUT charge of this further truncated sector that is relevant for us is, in Jacobi--Carter coordinates,
\bea
&& \ud s^2 = H \bigg[ \frac{r^2 + y^2}{R} \ud r^2 + \frac{r^2 +
  y^2}{Y} \ud y^2 - \frac{R}{H^2 (r^2 + y^2)} \mathcal{A}^2 \nnr
&& \qquad \qquad + \frac{Y}{r^2 + y^2} \left( \ud t' - r^2 \ud \psi_1 -
  \frac{q r}{H (r^2 + y^2)} \mathcal{A} \right) ^2 \bigg] , \nnr
&& X = H^{-1/2} , \quad A_{(1)}^1 = 0 , \quad A_{(1)}^2 = \frac{2 m s
  c r}{H (r^2 + y^2)} \mathcal{A} , \quad \chi = \frac{q y}{r^2 + y^2}
,
\eea
where
\bea
&& R = (1 + g^2 r^2) (r^2 + a^2) + q g^2 r (2 r^2 + a^2) + q^2 g^2 r^2 - 2mr , \nnr
&& Y = (1 - g^2 y^2) (a^2 - y^2) , \quad \mathcal{A} = \ud t + y^2 \ud \psi_1 , \nnr
&& H = 1 + \frac{q r}{r^2 + y^2} , \quad q = 2 m s^2 , \quad s = \sinh \delta , \quad c = \cosh \delta .
\eea
Taking $g = 0$, we recover the solution of (\ref{evengeneralPleb}) for $D = 4$.  Viewed as a solution of the $\textrm{U}(1)^4$ supergravity theory, in the notation of \cite{crotbh4d}, we have taken the four charge parameters to be $\delta_1 = \delta_3 = 0 ,~\delta_2 = \delta_4 = \delta$.

\subsection{Five dimensional gauged supergravity}

Performing a Kaluza--Klein reduction of type IIB supergravity on $S^5$
leads to $D = 5$, $\mathcal{N} = 8$,
$\textrm{SO}(6) \cong \textrm{SU}(4)$ gauged supergravity.  We
truncate to include only gauge fields in the $\textrm{U}(1)^3$ Cartan
subgroup of the full gauge group.  There is a consistent truncation
to minimal $\mathcal{N} = 2$ gauged supergravity coupled to two vector
multiplets.   The bosonic fields are a graviton, three $\textrm{U}(1)$
gauge fields and two scalars.

The Lagrangian is
\ben
\mathcal{L}_5 = R \star 1 - \frac{1}{2} \sum_{i=1}^2 \star \ud \varphi_i
\wedge \ud \varphi_i - \frac{1}{2} \sum_{I=1}^3 X_I^{-2} \star
F_{(2)}^I \wedge F_{(2)}^I + 4 g^2 \sum_{I=1}^3 X_I^{-1} \star 1 + F_{(2)}^1
\wedge F_{(2)}^2 \wedge A_{(1)}^3 ,
\een
where
\ben
X_1 = \ue ^{- \varphi_1 / \sqrt{6} - \varphi_2 / \sqrt{2}} , \quad X_2 =
\ue ^{- \varphi_1 / \sqrt{6} + \varphi_2 / \sqrt{2}} , \quad X_3 = \ue ^{2
  \varphi_1 / \sqrt{6}} , \quad F_{(2)}^I = \ud A_{(1)}^I .
\een

We may perform a consistent bosonic truncation to the sector with
$A_{(1)}^1 = A_{(1)}^2$ and $X = X_1 = X_2 = X_3^{-1/2} =
\ue ^{-\varphi_1 / \sqrt{6}}$.  Relabelling $\varphi_1 \rightarrow
\varphi$, the bosonic field equations can be obtained from the
Lagrangian
\bea
\mathcal{L}_5 & = & R \star 1 - \frac{1}{2} \star \ud \varphi \wedge \ud
\varphi - X^{-2} \star F_{(2)}^1 \wedge F_{(2)}^1 - \frac{1}{2} X^4
\star F_{(2)}^3 \wedge F_{(2)}^3 \nnr
&& + 4 g^2 (2 X^{-1} + X^2) \star 1 + F_{(2)}^1 \wedge F_{(2)}^1 \wedge A_{(1)}^3 .
\eea
Setting $g = 0$, we recover the Lagrangian of (\ref{dualhetL}).

For our purposes, we consider the black hole solution of
\cite{5dgsugrabhind} that has two independent rotation parameters, two
charges set equal, and the third charge set to a particular value once
the other charges are fixed, so there is one independent charge
parameter.  The solution in Jacobi--Carter coordinates is
\bea
&& \ud s^2 = H^{2/3} \bigg[ \frac{r^2 + y^2}{R} \ud r^2 + \frac{r^2 +
  y^2}{Y} \ud y^2 - \frac{R}{H^2 (r^2 + y^2)} \mathcal{A}^2 \nnr
&& \qquad \qquad \quad + \frac{Y}{r^2 + y^2} \left( \ud t' - r^2 \ud \psi_1 -
  \frac{q}{H (r^2 + y^2)} \mathcal{A} \right) ^2 \nnr
&& \qquad \qquad \quad + \frac{a^2 b^2}{r^2 y^2} \left( \ud t'+ (y^2 - r^2) \ud
  \psi_1 - r^2 y^2 \ud \psi_2 - \frac{q}{H (r^2 + y^2)} \mathcal{A}
\right) ^2 \bigg] , \nnr
&& X_1 = X_2 = H^{-1/3} , \quad X_3 = H^{2/3} , \nnr
&& A_{(1)}^1 = A_{(1)}^2 = \frac{2msc}{H (r^2 + y^2)} \mathcal{A}
, \quad A_{(1)}^3 = \frac{q ab}{r^2 + y^2} ( \ud \psi_1 + y^2 \ud
\psi_2 ) ,
\eea
where
\bea
&& R = \frac{(1 + g^2 r^2) (r^2 + a^2) (r^2 + b^2)}{r^2} + q g^2 (2 r^2 + a^2 + b^2) + q^2 g^2 - 2m , \nnr
&& Y = - \frac{(1 - g^2 y^2) (a^2 - y^2) (b^2 - y^2)}{y^2} , \quad \mathcal{A} = \ud t' + y^2 \ud \psi_1 , \nnr
&& H = 1 + \frac{q}{r^2 + y^2} , \quad q = 2 m s^2 , \quad s = \sinh \delta , \quad c = \cosh \delta .
\eea
Taking $g = 0$, we recover the solution of (\ref{oddgeneralPleb}) for $D = 5$.

A generalization of this solution has recently been discovered such that
although there are two equal $\textrm{U}(1)$ charges, the third may be
independently specified \cite{newrotbh5d}; this generalization also
includes the five dimensional minimal gauged supergravity black hole
solution of \cite{bh5dgsugra}.  However, as this more general
solution has two independent charge parameters rather than one, we do
not require any of its additional features to guide us towards the new
seven dimensional solution obtained in the next section.

\section{Seven dimensional gauged supergravity black holes}

Reducing eleven dimensional supergravity on $S^4$ leads to $D = 7$,
$\mathcal{N} = 4$, $\textrm{SO}(5)$ gauged supergravity.  We truncate
to include only gauge fields in the $\textrm{U}(1)^2$ Cartan subgroup of the full
gauge group.  The bosonic fields are a graviton, a three-form
potential, two $\textrm{U}(1)$ gauge fields and two scalars.

The bosonic Lagrangian is
\bea
\mathcal{L}_7 & = & R \star 1 - \frac{1}{2} \sum_{i=1}^2 \star \ud \varphi_i \wedge
\ud \varphi_i - \frac{1}{2} \sum_{I=1}^2 X_I^{-2} \star F_{(2)}^I
\wedge F_{(2)}^I - \frac{1}{2} X_1^2 X_2^2 \star F_{(4)} \wedge
F_{(4)} \nnr
&& + 2 g^2 (8 X_1 X_2 + 4 X_1^{-1} X_2^{-2} + 4 X_1^{-2} X_2^{-1} -
X_1^{-4} X_2^{-4}) \star 1 \nnr
&& + g F_{(4)} \wedge A_{(3)} + F_{(2)}^1 \wedge F_{(2)}^2 \wedge
A_{(3)} ,
\eea
where
\ben
X_1 = \ue ^{- \varphi_1 / \sqrt{10} - \varphi_2 / \sqrt{2}} , \quad X_2 =
\ue ^{ - \varphi_1 / \sqrt{10} + \varphi_2 / \sqrt{2}} , \quad F_{(2)}^I =
\ud A_{(1)}^I , \quad F_{(4)} = \ud A_{(3)} .
\een
The resulting Einstein equation is
\bea
G_{ab} & = & \sum_{i=1}^2 \left( \frac{1}{2} \nabla_a \varphi_i
\nabla_b \varphi_i - \frac{1}{4} \nabla^c \varphi_i \nabla_c \varphi_i
g_{ab} \right) + \sum_{I=1}^2 X_I^{-2} \left( \frac{1}{2} F{^I}{_{a}}{^c}
  F{^I}{_{bc}} - \frac{1}{8} F^{Icd} F{^I}{_{cd}} g_{ab} \right) \nnr
&& + X_1^2 X_2^2 \left( \frac{1}{12} F{_a}{^{cde}} F_{bcde} -
  \frac{1}{96} F^{cdef} F_{cdef} g_{ab} \right) \nnr
&& + g^2 (8 X_1 X_2 + 4 X_1^{-1} X_2^{-2} + 4
X_1^{-2} X_2^{-1} - X_1^{-4} X_2^{-4}) g_{ab} .
\eea
The remaining field equations are
\bea
&& \square \varphi_1 = \frac{1}{2 \sqrt{10}} \sum_{I=1}^2 X_I^{-2} F^{Iab}
F{^I}{_{ab}} - \frac{1}{12 \sqrt{10}}
X_1^2 X_2^2 F^{abcd} F_{abcd} , \nnr
&& \qquad \quad + \frac{8}{\sqrt{10}} g^2 (4 X_1 X_2 - 3
X_1^{-1} X_2^{-2} - 3 X_1^{-2} X_2^{-1} + 2 X_1^{-4} X_2^{-4}) , \nnr
&& \square \varphi_2 = \frac{1}{2 \sqrt{2}} (X_1^{-2} F^{1ab}
F{^1}{_{ab}} - X_2^{-2} F^{2ab} F{^2}{_{ab}}) + 4 \sqrt{2} g^2
(X_1^{-1} X_2^{-2} - X_1^{-2} X_2^{-1}) , \nnr
&& \ud (X_1^{-2} \star F_{(2)}^1) = F_{(2)}^2 \wedge F_{(4)} , \nnr
&& \ud (X_2^{-2} \star F_{(2)}^2) = F_{(2)}^1 \wedge F_{(4)} , \nnr
&& \ud (X_1^2 X_2^2 \star F_{(4)}) = 2 g F_{(4)} + F_{(2)}^1 \wedge
F_{(2)}^2 .
\eea
Once the field equations arising from the Lagrangian are satisfied,
there is also a self-duality condition to impose, which can be stated
by including a two-form potential $A_{(2)}$ and defining
\ben
F_{(3)} = \ud A_{(2)} - \frac{1}{2} A_{(1)}^1 \wedge \ud A_{(1)}^2 -
\frac{1}{2} A_{(1)}^2 \wedge \ud A_{(1)}^1 .
\een
The self-duality equation is
\ben
X_1^2 X_2^2 \star F_{(4)} = 2 g A_{(3)} - F_{(3)} .
\een

If we truncate to solutions with $X = X_1 = X_2 = \ue ^{- \varphi_1 /
  \sqrt{10}}$, $\varphi_2 = 0$ and $A_{(1)} = A_{(1)}^1 = A_{(1)}^2$,
then the bosonic field equations can be obtained from the Lagrangian
\bea
\mathcal{L}_7 & = & R \star 1 - \frac{1}{2} \star \ud \varphi_1 \wedge
  \ud \varphi_1 - X^{-2} \star F_{(2)} \wedge F_{(2)} - \frac{1}{2}
  X^4 \star F_{(4)} \wedge F_{(4)} \nnr
&& + 2 g^2 (8 X^2 + 8 X^{-3} - X^{-8}) \star 1 + F_{(2)} \wedge
  F_{(2)} \wedge A_{(3)} - g F_{(4)} \wedge A_{(3)} ,
\eea
where $F_{(2)} = \ud A_{(1)}$.  Setting $g = 0$, we recover the
  Lagrangian of (\ref{dualhetL}).  The self-duality equation becomes
\ben
X^4 \star F_{(4)} = 2 g A_{(3)} - \ud A_{(2)} + F_{(2)} \wedge A_{(1)} .
\een

\subsection{Black hole solutions}

Before presenting the new black hole solution, we first review the
known black hole solutions in seven dimensions.  The starting point
for rotating black holes in higher dimensions is the Myers--Perry
black hole \cite{bhhigherdim}, which generalizes the Kerr solution of
four dimensions.  In arbitrary dimensions, there is a generalization
to include a cosmological constant \cite{genKerrdS, rotbhhigherdim},
and, as discussed previously, in the context of toroidally
compactified supergravity, a generalization to include charges
\cite{nearBPSsat}.  In seven dimensions, viewed in the context of
$\textrm{U}(1)^2$ gauged supergravity, these respectively provide
black hole solutions with three arbitrary rotation parameters, no
charges, and arbitrary gauge coupling constant $g$; and with three
arbitrary rotation parameters, two arbitrary charges, and zero gauge
coupling.

Specific to seven dimensional gauged supergravity, a non-rotating
black hole with two independent $\textrm{U}(1)$ charges and arbitrary
gauge coupling constant was given in \cite{phasesRcbh, bhmemAdS7}.  A
generalization to include three equal angular momenta, as well as the
two independent $\textrm{U}(1)$ charges and arbitrary gauge coupling,
was obtained in \cite{crotbh7d}.  The solution involves the
Fubini--Study metric on $\mathbb{CP}^2$,
\ben
\ud \Sigma_2^2 = \ud \xi^2 + \frac{1}{4} \sin^2 \xi (\sigma_1^2 +
\sigma_2^2) + \frac{1}{4} \sin^2 \xi \cos^2 \xi \sigma_3^2 ,
\een
where $\sigma_i$ are left-invariant one-forms on $\textrm{SU}(2)$
that satisfy $\ud \sigma_i = - \frac{1}{2} \epsilon_{ijk} \sigma_j
\wedge \sigma_k$, for which the K\"{a}hler form is $J = \frac{1}{2}
\ud \sigma$ with $\sigma = \ud \tau + \frac{1}{2} \sin^2 \xi
\sigma_3$.  A simplification occurs if the two $\textrm{U}(1)$ charges
are set equal.  Compared with \cite{crotbh7d}, we perform the
coordinate changes $t \rightarrow (1 - ag) t$, followed by $\tau \rightarrow \tau - g t$, so $\sigma
\rightarrow \sigma - g \ud t$, and so the solution is written as
\bea
&& \ud s^2 = H^{2/5} \bigg[ - \frac{V - 2m}{H^2 \Xi^2 (r^2 + a^2)^2} (\ud t
- a \sigma)^2 + \frac{(r^2 + a^2)^2}{V - 2m} \ud r^2 + \frac{r^2 + a^2}{\Xi} \ud
\Sigma_2^2 \nnr
&& \qquad \qquad \quad + \frac{a^2}{\Xi^2 r^2} \left( (1 + g^2 r^2) \ud t - \frac{r^2 + a^2}{a} \sigma - \frac{2 m s^2 (1 + ag)}{H
    (r^2 + a^2)^2} (\ud t - a \sigma) \right) ^2 \bigg] , \nnr
&& X = H^{-1/5} , \quad A_{(1)} = \frac{2msc}{H \Xi (r^2 + a^2)^2} (\ud t - a
\sigma) , \quad A_{(2)} = \frac{2 m s^2 (1 + ag)}{H \Xi (r^2 + a^2)^2} \ud t
\wedge a \sigma , \nnr
&& A_{(3)} = \frac{2 m s^2}{\Xi^2 (r^2 + a^2)^2} [a (\sigma - a g^2 \ud t) - ag (\ud t
- a \sigma)] \wedge (r^2 + a^2) J ,
\eea
where
\bea
&& V = \frac{(1 + g^2 r^2) (r^2 + a^2)^3}{r^2} + 2 m s^2 g^2 (2 r^2 +
3 a^2) - \frac{4 m s^2 g a^3}{r^2} + \frac{(2 m s^2)^2 g^2}{r^2} ,
\nnr
&& H = 1 + \frac{2 m
  s^2}{(r^2 + a^2)^2} , \quad s = \sinh \delta , \quad c = \cosh \delta , \quad \Xi = 1 - a^2 g^2 .
\eea
The time coordinate $t$ that appears here is canonically normalized and matches the $t$ of, for example, \cite{genKNUTAdSalld}.

Guided by the structure of the solutions we have just discussed, we
are in a position to obtain a new seven dimensional solution with
three independent angular momenta and equal charges.  In particular,
we are helped by the simple form of the sevenbeins in terms of which
the previously known solutions may be written.

The new solution is
\bea
&& \ud s^2 = H^{2/5} \bigg\{ \frac{(r^2 + y^2) (r^2 + z^2)}{R} \ud r
^2 + \frac{(r^2 + y^2) (y^2 - z^2)}{Y} \ud y^2 + \frac{(r^2 + z^2)
  (z^2 - y^2)}{Z} \ud z^2 \nnr
&& \quad \qquad - \frac{R}{H^2 (r^2 + y^2) (r^2 + z^2)} \mathcal{A}^2 \nnr
&& \quad \qquad + \frac{Y}{(r^2 + y^2) (y^2 - z^2)} \left[ \ud t' +
  (z^2 - r^2) \ud \psi_1 - r^2 z^2 \ud \psi_2 - \frac{q}{H (r^2 + y^2)
    (r^2 + z^2)} \mathcal{A} \right] ^2 \nnr
&& \quad \qquad + \frac{Z}{(r^2 + z^2) (z^2 - y^2)} \left[ \ud t' +
  (y^2 - r^2) \ud \psi_1 - r^2 y^2 \ud \psi_2 - \frac{q}{H (r^2 + y^2)
    (r^2 + z^2)} \mathcal{A} \right] ^2 \nnr
&& \quad \qquad + \frac{a_1^2 a_2^2 a_3^2}{r^2 y^2 z^2} \bigg[ \ud t'
+ (y^2 + z^2 - r^2) \ud \psi_1 + (y^2 z^2 - r^2 y^2 - r^2 z^2) \ud
\psi_2 - r^2 y^2 z^2 \ud \psi_3 \nnr
&& \quad \qquad \qquad \qquad - \frac{q}{H (r^2 + y^2) (r^2 +
  z^2)} \left( 1 + \frac{g y^2 z^2}{a_1 a_2 a_3} \right) \mathcal{A} \bigg] ^2 \bigg\} , \nnr
&& X = H^{-1/5} , \quad A_{(1)} = \frac{2msc}{H (r^2 + y^2) (r^2 + z^2)} \mathcal{A} , \nnr
&& A_{(3)} = q a_1 a_2 a_3 [\ud \psi_1 + (y^2 + z^2) \ud \psi_2 + y^2 z^2 \ud \psi_3] \nnr
&& \qquad \quad \wedge \left( \frac{1}{(r^2 + y^2) z} \ud z \wedge
  (\ud \psi_1 + y^2 \ud \psi_2) + \frac{1}{(r^2 + z^2) y} \ud y \wedge
  (\ud \psi_1 + z^2 \ud \psi_2) \right) \nnr
&& \qquad \quad - q g \mathcal{A} \wedge \left( \frac{z}{r^2 + y^2}
  \ud z \wedge (\ud \psi_1 + y^2 \ud \psi_2) + \frac{y}{r^2 + z^2} \ud
  y \wedge (\ud \psi_1 + z^2 \ud \psi_2) \right) ,
\label{7dsolmetric}
\eea
where
\bea
&& R = \frac{1 + g^2 r^2}{r^2} \prod_{i=1}^3 (r^2 + a_i^2) + q
g^2 ( 2 r^2 + a_1^2 + a_2^2 + a_3^2 ) - \frac{2 q g a_1 a_2 a_3}{r^2} +
\frac{q^2 g^2}{r^2} - 2m , \nnr
&& Y = \frac{1 - g^2 y^2}{y^2} \prod_{i=1}^3 (a_i^2 - y^2) , \quad Z =
\frac{1 - g^2 z^2}{z^2} \prod_{i=1}^3 (a_i^2 - z^2) , \nnr
&& \mathcal{A} = \ud t' + (y^2 + z^2) \ud \psi_1 + y^2 z^2 \ud \psi_2 , \nnr
&& H = 1 + \frac{q}{(r^2 + y^2) (r^2 + z^2)} , \quad q = 2 m s^2 , \quad s = \sinh \delta , \quad c = \cosh \delta .
\eea
The two-form potential is
\bea
A_{(2)} & = & \frac{q}{H (r^2 + y^2) (r^2+z^2)} \mathcal{A} \wedge \nnr
&& \bigg( \ud t' + \sum_i a_i^2 (g^2 \ud t' + \ud \psi_1) + \sum_{i <
  j} a_i^2 a_j^2 (g^2 \ud \psi_1 + \ud \psi_2) + a_1^2 a_2^2 a_3^2
(g^2 \ud \psi_2 + \ud \psi_3) \nnr
&& ~ - g^2 (y^2 + z^2) \ud t' - g^2 y^2 z^2 \ud \psi_1 + a_1 a_2 a_3 g
[\ud \psi_1 + (y^2 + z^2) \ud \psi_2 + y^2 z^2 \ud \psi_3] \bigg) .
\eea
It is straightforward to verify on a computer that the above solution
does indeed satisfy the field equations.  The natural choice of
sevenbeins facilitates the computations, including those of the metric
determinant and of the metric inverse.

The structure of this seven dimensional solution is analogous
to the five dimensional gauged supergravity solution of \cite{5dgsugrabhind} with two equal $\textrm{U}(1)$
charges and the third $\textrm{U}(1)$ charge equal to a particular
value.  However, the solution could also be thought of as
analogous to the five dimensional black hole of minimal gauged
supergravity, for which all three $\textrm{U}(1)$ charges are set
equal \cite{bh5dgsugra}.

For computing thermodynamical quantities, as advocated in
\cite{1stlaw}, we use angular velocities measured with respect to a
non-rotating frame at infinity and move to Boyer--Lindquist
coordinates using the coordinate change \cite{genKNUTAdSalld}
\bea
t & = & t' + (a_1^2 + a_2^2 + a_3^2) \psi_1 + (a_1^2 a_2^2 + a_2^2
a_3^2 + a_3^2 a_1^2) \psi_2 + a_1^2 a_2^2 a_3^2 \psi_3 , \nnr
\frac{\phi_i}{a_i} & = & g^2 t' + \psi_1 + \sum_{j \neq i} a_j^2 (g^2
\psi_1 + \psi_2) + \prod_{j \neq i} a_j^2 (g^2 \psi_2 + \psi_3) .
\eea
It is routine to perform the coordinate change, although it may be
helpful to note here that the two-form potential is
\bea
A_{(2)} & = & \frac{q (a_1 + a_2 a_3 g)}{H (r^2 + y^2) (r^2 + z^2)} \nnr
&& \left( \frac{(1 - g^2 y^2) (1 - g^2 z^2)}{\Xi_1 \Xi_2 \Xi_3} \ud t
  \wedge \mu_1^2 \ud \phi_1 + \frac{g (a_3^2 - a_2^2) \mu_2^2
    \mu_3^2}{\Xi_2 \Xi_3} \ud \phi_2 \wedge \ud \phi_3 \right) +
\textrm{cyclic} ,
\eea
where there are two additional terms by cycling the indices $1, 2, 3$, and, from (\ref{mui}),
\ben
\mu_i^2 = \frac{(a_i^2 - y^2) (a_i^2 - z^2)}{\prod_{j \neq i} (a_i^2 - a_j^2)} .
\een
We have also denoted $\Xi_i = 1 - a_i^2 g^2$, which are positive so that the signature is correct.  In these $(t, r, y, z, \phi_1,
\phi_2, \phi_3)$ coordinates, the metric determinant is
\ben
\det (g_{ab}) = \frac{H^{4/5} r^2 y^2 z^2 (r^2 + y^2)^2 (r^2 +
  z^2)^2 (y^2-z^2)^2}{(a_1^2-a_2^2)^2 (a_2^2-a_3^2)^2 (a_3^2-a_1^2)^2} .
\label{BLdet}
\een

\subsection{Curvature singularities}

The presentation of the higher dimensional Kerr--NUT--AdS solution of \cite{genKNUTAdSalld} gives a simple orthonormal basis with which one may compute the curvature, as done in \cite{KNdScurv}.  The structure of the curvature singularities is similar to that discussed in \cite{bhhigherdim} for the Myers--Perry solution.  If any rotation parameter $a_i$ vanishes, then there is a curvature singularity at $r = 0$.  For the general case in which none of the rotation parameters $a_i$ vanish, we might be worried about singular behaviour at $r = 0$ since there are singularities in the vielbein components there.  However, it turns out that all the curvature components $R_{\mu \nu \rho \sigma}$ in this orthonormal basis are well-behaved at $r = 0$ provided no $a_i$ vanishes, and also the metric components $g_{ab}$ and inverse metric components $g^{ab}$ are well-behaved there.

For the seven dimensional solution (\ref{7dsolmetric}) we have obtained, we may perform the coordinate change $u = r^2$, and it turns out that the metric components $g_{ab}$ and inverse metric components $g^{ab}$, as well as the metric determinant $\det(g_{ab})$, are all non-singular at $u = 0$.  A simple choice of orthonormal frame may be read of from the way the metric is presented, and includes
\bea
e_0 & = & H^{-4/5} \frac{\sqrt{R}}{\sqrt{(r^2 + y^2) (r^2 + z^2)}} \mathcal{A} , \nnr
e_6 & = & \frac{H^{1/5} a_1 a_2 a_3}{r y z} \bigg[ \ud t'
+ (y^2 + z^2 - r^2) \ud \psi_1 + (y^2 z^2 - r^2 y^2 - r^2 z^2) \ud
\psi_2 - r^2 y^2 z^2 \ud \psi_3 \nnr
&& \qquad \qquad \qquad - \frac{q}{H (r^2 + y^2) (r^2 +
  z^2)} \left( 1 + \frac{g y^2 z^2}{a_1 a_2 a_3} \right) \mathcal{A} \bigg] .
\eea
However, in the $(t', r, y, z, \psi_1, \psi_2, \psi_3)$ coordinates, regardless of any coordinate change $ u = r^2$, there remains singular behaviour in the sevenbein components $e_0$ and $e_6$ at $r = 0$.  These singularities conspired to cancel each other out when forming metric components, but one might still be wary of a curvature singularity there, as some orthonormal components of the curvature $R_{\mu \nu \rho \sigma}$ diverge.  However, such an apparent singularity is caused by a bad choice of orthonormal frame.  We have, in these coordinates, orthonormal frame components
\bea
e_0 & = & \left( 1 + \frac{q}{y^2 z^2} \right) ^{-4/5} \frac{a_1 a_2 a_3 - qg}{ryz} \mathcal{A} + \textrm{O}(r) , \nnr
e_6 & = & \left( 1 + \frac{q}{y^2 z^2} \right) ^{-4/5} \frac{a_1 a_2 a_3 - qg}{ryz} \mathcal{A} + \textrm{O}(r) ,
\eea
so there is a degeneracy in the orthonormal frame as $r \rightarrow 0$.  We may remove the degeneracy by changing orthonormal frames and performing local Lorentz boosts with arbitrarily large rapidity as $r \rightarrow 0$.  For example, we could replace $e_0$ and $e_6$ in favour of
\bea
e' {_0} & = & \frac{1}{r} e_0 - \frac{\sqrt{1 - r^2}}{r} e_6 = \textrm{O}(r^0) , \nnr
e' {_6} & = & \frac{1}{r} e_6 - \frac{\sqrt{1 - r^2}}{r} e_0 = \textrm{O}(r^0) ,
\eea
leaving $e_1 , \ldots , e_5$ unchanged.  The new inverse sevenbein components $e^a_\mu$ are also well-behaved at $r = 0$.  It follows that the curvature components $R_{\mu \nu \rho \sigma}$ in this orthonormal frame must be non-singular at $r = 0$, so the geometry is regular there.  Using the new radial coordinate $u = r^2$, the metric may be extended to negative values of $u$, which may be thought of as extending to imaginary values of $r$.  If one examines the orthonormal components of the Riemann tensor, then one finds negative powers of $r^2 + y^2$ and $r^2 + z^2$, and so there is a curvature singularity that extends out to a maximum radius given by $r^2 = - \min (a_1^2, a_2^2, a_3^2)$.  There are also negative powers of $(r^2 + y^2) (r^2 + z^2) + q$ appearing in the curvature, which moves the curvature singularity further out for $q < 0$.  If there is a horizon at $r = r_0$ or some minimum radius to the geometry at $r = r_0$, then a naked singularity is avoided if $r_0^2 + a_i^2 > 0$ for each $i$ and $(r_0^2 + a_i^2) (r_0^2 + a_j^2) + q > 0$ for each $i \neq j$.  From now on, we assume that the parameters of the solution are chosen so that the outermost curvature singularity is hidden behind a horizon.

\subsection{Thermodynamics}

The outer black hole horizon is located at the largest root of $R(r)$, say
at $r = r_+$.  Its angular velocities are constant over the horizon
and are obtained from the Killing vector $\ell = \pd / \pd t + \sum_i
\Omega_i \pd / \pd \phi_i$ that becomes null on the horizon.  The
surface gravity $\kappa$, also constant over the horizon, is given by
$\ell^b \nabla_b \ell^a = \kappa \ell^a$ evaluated on the horizon.
The horizon area is obtained from integrating the square root
of the determinant of the induced metric on a time slice of the
horizon,
\ben
\det g_{(y, z, \phi_1, \phi_2, \phi_3)} |_{r = r_+} = \frac{[ \prod_i (r_+^2 + a_i^2) + q (r_+^2 - a_1 a_2 a_3 g)]^2 y^2 z^2 (y^2 -
  z^2)^2}{\Xi_1^2 \Xi_2^2 \Xi_3^2 (a_1^2 - a_2^2)^2 (a_2^2 - a_3^2)^2
  (a_3^2 - a_1^2)^2 r_+^2} .
\label{areadet}
\een
Bearing in mind that the radial coordinate may be analytically continued to negative values of $r^2$, we should demand that $r_+^2 > 0$, so the horizon area, or equivalently the entropy, is real.  We take the temperature to be $T = \kappa / 2 \pi$ and the entropy to
be one quarter of the horizon area.  The angular momenta are given by
the Komar integrals
\ben
J_i = \frac{1}{16 \pi} \int_{S^5} \star \ud K_i ,
\een
where $K_i$ is the one-form obtained from the Killing vector $\pd /
\pd \phi_i$.  Using the Killing vector $\ell$ that becomes null on the
horizon, we obtain the electrostatic potential $\Phi = \ell \cdot A_{(1)}$ evaluated on the horizon, over which it is constant.  The
conserved Page electric charge is
\ben
Q = \frac{1}{8 \pi} \int_{S^5} ( X^{-2} \star F_{(2)} - F_{(2)} \wedge A_{(3)} ) ,
\een
although for our solution there is no contribution from the $F_{(2)}
\wedge A_{(3)}$ term; our normalization factor of $1 / 8 \pi$ rather
than $1 / 16 \pi$ arises from using the canonical normalization for
two separate $\textrm{U}(1)$ fields and then setting them equal.

One finds that $T \ud S + \sum_i \Omega_i \ud J_i + \Phi \ud Q$ is an
exact differential, and so we may integrate the first law of black
hole mechanics,
\ben
\ud E = T \ud S + \sum_i \Omega_i \ud J_i + \Phi \ud Q ,
\een
to obtain an expression for the thermodynamic mass $E$.  There are
various other methods of obtaining the energy of an asymptotically AdS
spacetime that we do not pursue here, but discussion of how various
methods are applied to computing the conserved charges of AdS black
hole solutions may be found in \cite{1stlaw, massrotbh} for example.

In summary, we find the thermodynamical quantities
\bea
E & = & \frac{\pi^2}{8 \Xi_1 \Xi_2 \Xi_3} \left[ \sum_i \frac{2m}{\Xi_i}
  - m + \frac{5q}{2} + \frac{q}{2} \sum_i
  \left(  \sum_{j
      \neq i} \frac{2 \Xi_j}{\Xi_i} - \Xi_i - \frac{2 (1 + 2 a_1 a_2 a_3 g^3)}{\Xi_i} \right) \right] , \nnr
T & = & \frac{(1 + g^2 r_+^2) r_+^2 \sum_i \prod_{j \neq i} (r_+^2 + a_j^2)
  - \prod_i (r_+^2 + a_i^2)  + 2q ( g^2 r_+^4 + g a_1 a_2 a_3 ) - q^2 g^2}{2
  \pi r_+ [(r_+^2 + a_1^2) (r_+^2 + a_2^2) (r_+^2 +
  a_3^2) + q (r_+^2 - a_1 a_2 a_3 g)]}, \nnr
S & = & \frac{\pi^3 [(r_+^2 + a_1^2) (r_+^2 + a_2^2) (r_+^2 + a_3^2) + q (r_+^2 - a_1 a_2 a_3 g)]}{4 \Xi_1 \Xi_2 \Xi_3 r_+} , \nnr
\Omega_i & = & \frac{a_i [(1 + g^2 r_+^2) \prod_{j \neq i} (r_+^2 +
  a_j^2) + q g^2 r_+^2] - q \prod_{j \neq i} a_j g}{(r_+^2
  + a_1^2) (r_+^2 + a_2^2) (r_+^2 + a_3^2) + q (r_+^2 - a_1 a_2
  a_3 g)} , \nnr
J_i & = & \frac{\pi^2 m [a_i c^2 - s^2 g (\prod_{j \neq i} a_j + a_i
  \sum_{j \neq i} a_j^2 g + a_1 a_2 a_3 a_i g^2) ]}{4 \Xi_1 \Xi_2 \Xi_3 \Xi_i} , \nnr
\Phi & = & \frac{2msc r_+^2}{(r_+^2 + a_1^2) (r_+^2 + a_2^2) (r_+^2 + a_3^2) + q (r_+^2 - a_1 a_2 a_3 g)} , \nnr
Q & = & \frac{\pi^2 m s c}{\Xi_1 \Xi_2 \Xi_3} .
\eea

The Gibbs free energy, $G = E - TS - \sum_i \Omega_i J_i - \Phi Q$, is
\bea
G & = & \frac{\pi^2}{16 \Xi_1 \Xi_2 \Xi_3 r_+^2} \Bigg\{ (1 - g^2 r_+^2)
\prod_i (r_+^2 + a_i^2) - 2 q g^2 r_+^4 - 2 q a_1 a_2 a_3 g \nnr
&& \qquad \qquad \qquad - q^2 \Bigg[ g^2 \left( \sum_i a_i^2 r_+^4 - \sum_{i < j} a_i^2 a_j^2
  r_+^2 - \prod_i a_i^2 \right) + a_1 a_2 a_3 g (2 g^2 r_+^4 - 2 r_+^2 + q g^2) \nnr
&& \qquad \qquad \qquad \qquad + g^2 r_+^2 (r_+^4 + q) \Bigg] \left( \prod_i (r_+^2 + a_i^2) + q (r_+^2 - a_1 a_2 a_3 g) \right) ^{-1} \Bigg\} .
\eea
One then obtains the grand canonical partition function, $Z_{\textrm{gc}} = \exp (-G/T)$.

For black holes with a large horizon radius compared to the AdS
radius, there are universal thermodynamical predictions arising from
conformal fluid mechanics via the AdS/CFT correspondence \cite{thermspinbranes,
  AdSbhfluid}.  We set $g = 1$ and take the limit $r_+ \gg 1$ keeping
$k = q /r_+^4$ fixed.  In this large black hole limit, after
multiplying integrals by $1 / g^5 G_7 = 16 N^3 / 3 \pi^2$, the
thermodynamics is summarized by
\ben
T = \frac{r_+ (3 - k)}{2 \pi} , \quad \Omega_i = a_i , \quad \Phi = 2
\pi T \frac{\sqrt{k}}{3 - k} , \quad \ln Z_{\textrm{gc}} = \frac{64
  \pi^6 N^3 T^5}{3 \prod_i (1 - \Omega_i^2)} \frac{(1+k)^2}{(3 - k)^6}
.
\een
These agree with the fluid mechanical predictions, with the first
corrections being $\textrm{O}(1 / r_+^2)$.

\subsection{BPS limit}

Conditions for a solution to be BPS were obtained in
\cite{rotbhgaugedsugra} from considering eigenvalues of the Bogomolny
matrix.  In our case, for which the two $\textrm{U}(1)$ charges have
been set equal, once appropriate signs of $\phi_i$ and $A_{(1)}$ have
been chosen, a BPS solution
satisfies
\ben
\quad E + g \sum_i J_i - Q = 0 .
\label{BPS}
\een
It should be noted that there is a typographical error in \cite{rotbhgaugedsugra} concerning conditions for the vanishing of the eigenvalues of the Bogomolny matrix.  Specifically, equation (4.14) should be $\ue ^{\delta_1 + \delta_2} = 1 + 2/ag, ~ (1+2/ag)^{-1}, ~ 1-2/3ag, ~ (1-2/3ag)^{-1}$, since $E$ and $J$ are invariant under $\delta_i \rightarrow - \delta_i$, but $Q_i \rightarrow - Q_i$; therefore $\ue ^{\delta_1 + \delta_2} = 1 + 2/ag$ corresponds to $E - gJ - \sum_i Q_i = 0$ and $\ue ^{\delta_1 + \delta_2} = 1 - 2/3ag$ corresponds to $E + 3gJ - \sum_i Q_i = 0$.

The BPS condition is satisfied if
\ben
\ue ^{2 \delta} = 1 - \frac{2}{(a_1 + a_2 + a_3) g} ,
\label{BPSdelta}
\een
which recovers the type A and type B conditions of
\cite{rotbhgaugedsugra} on setting the $a_i$ equal up to signs.
For $\delta$ to be real, we must have $(a_1 + a_2 + a_3) g < 0$ or $(a_1 + a_2 + a_3) g > 2$, along with the previous requirement that $-1 < a_i g < 1$.  Equivalently, in a form that is more directly useful, the BPS
constraint is
\ben
q = - \frac{2m}{(a_1 + a_2 + a_3) g (2 - a_1 g - a_2 g - a_3 g)} .
\een
The Killing vector
\ben
K = \frac{\pd}{\pd t} - g \sum_i \frac{\pd}{\pd \phi_i}
\een
is the square of a Killing spinor $\epsilon$, i.e.~$K^a =
\bar{\epsilon} \gamma^a \epsilon$.  Because of its spinorial square
root, from Fierz identities one may show that $K$ is non-spacelike, and we have $g(K, K) = -
f^2$, with
\ben
f = H^{-4/5} \left( 1 + \frac{ qg (1 - a_1 g - a_2 g - a_3 g)^2 (r_\text{h}^2 +
    y^2) (r_\text{h}^2 + z^2)}{\Xi_{1-} \Xi_{2-} \Xi_{3-} (a_1+a_2)
      (a_2+a_3) (a_3+a_1) (r^2 + y^2) (r^2 + z^2)} \right) ,
\een
where
\ben
r_\text{h}^2 = \frac{a_1 a_2 + a_2 a_3 + a_3 a_1 - a_1 a_2 a_3 g}{1 - a_1 g - a_2 g - a_3 g} .
\label{r1}
\een
We have also used the definition $\Xi_{i \pm} = 1 \pm a_i g$.

The supersymmetric solutions generally preserve $\frac{1}{8}$
supersymmetry, although there can be supersymmetry enhancement if more
than one eigenvalue of the Bogomolny matrix vanishes.  We should
recall that these eigenvalues are
\bea
\lambda_{i \pm} & = & E + g J_i - g \sum_{j \neq i} J_j \pm Q , \nnr
\lambda_{4 \pm} & = & E + g (J_1 + J_2 + J_3) \pm Q ;
\label{Bogevals}
\eea
the number of supersymmetries preserved is the number of zero eigenvalues.  We have chosen conventions (\ref{BPS}) such that
$\lambda_{4-} = 0$, but it may be possible for some of the other
eigenvalues to also vanish.  Since the charge $Q$ does not vanish for
the BPS solutions, as follows from (\ref{BPSdelta}), apart from for
$\textrm{AdS}_7$ itself, we see that at most four of the eight eigenvalues (\ref{Bogevals}) of the
Bogomolny matrix can vanish.  The possibilities for
enhanced supersymmetry are, up to permutations,
\ben
\setlength\arraycolsep{5pt}
\begin{array}{l l l}
\frac{1}{4} \textrm{ supersymmetric} & \frac{3}{8} \textrm{ supersymmetric} & \frac{1}{2} \textrm{ supersymmetric} \\
\lambda_{1+} = 0 & \lambda_{1+} = \lambda_{2+} = 0 & \lambda_{1+} = \lambda_{2+} = \lambda_{3+} = 0 \\
\lambda_{1-} = 0 & \lambda_{1+} = \lambda_{2-} = 0 & \lambda_{1+} = \lambda_{2+} = \lambda_{3-} = 0 \\
& \lambda_{1-} = \lambda_{2-} = 0 & \lambda_{1+} = \lambda_{2-} = \lambda_{3-} = 0 \\
& & \lambda_{1-} = \lambda_{2-} = \lambda_{3-} = 0 \\
\end{array}
.
\label{moresusy}
\een
The supersymmetric solutions might therefore be $\frac{1}{8}, \frac{1}{4},
\frac{3}{8}, \frac{1}{2}$ supersymmetric.  Given a BPS solution, the eigenvalue $\lambda_{1+}$ vanishes if
\ben
a_1 g = \frac{4 - (a_2 + a_3) g - 3 (a_2^2 + a_3^2) g^2 - 2 a_2 a_3
  g^2 - a_2 a_3 (a_2 + a_3) g^3}{4 + (a_2 + a_3) g - (a_2^2 + a_3^2)
  g^2 + 2 a_2 a_3 g^2 + a_2 a_3 (a_2 + a_3) g^3} .
\label{lambda1+}
\een
The eigenvalue $\lambda_{1-}$ vanishes if either of the following two conditions holds:
\bea
&& a_1 g = \frac{1 - (a_2 + a_3) g - 3 a_2 a_3 g^2}{3 + (a_2 + a_3) g - a_2 a_3 g^2} , \label{lambda1-a} \\
&& a_2 + a_3 = 0 \label{lambda1-b} .
\eea
However, we must take into
account the inequalities that the rotation parameters must satisfy:
$-1 < a_i g < 1$ for the correct signature; from the BPS constraint (\ref{BPSdelta}), either $(a_1 + a_2
+ a_3) g < 0$ or $(a_1 + a_2 + a_3) g > 2$; for a real horizon area
and entropy, a horizon $r = r_0$ must have $r_0^2 > 0$; and conditions
to avoid a naked singularity.

To investigate whether or not the spacetime suffers from the pathology
of naked closed timelike curves (CTCs), we write the metric in the
form
\bea
\ud s^2 & = & H^{2/5} \bigg[ \frac{(r^2 + y^2) (r^2 + z^2)}{R} \ud r
^2 + \frac{(r^2 + y^2) (y^2 - z^2)}{Y} \ud y^2 + \frac{(r^2 + z^2)
  (z^2 - y^2)}{Z} \ud z^2 \nnr
&& \qquad - \frac{r^2 y^2 z^2 RYZ}{H^2 \prod_{i < j} (a_i^2 - a_j^2)^2
  B_1 B_2 B_3} \ud t^2 + B_3 (\ud \phi_3 + v_{32} \ud \phi_2 + v_{31}
\ud \phi_1 + v_{30} \ud t)^2 \nnr
&& \qquad + B_2 (\ud \phi_2 + v_{21} \ud \phi_1 + v_{20} \ud t)^2 +
B_1 (\ud \phi_1 + v_{10} \ud t)^2 \bigg] ,
\eea
so that the periodic $\phi_i$ coordinates have been separated from a
$\ud t ^2$ term.  We have used (\ref{BLdet}) in writing this form of
the metric, but for now do not require any details of the additional
functions introduced,
which one can straightforwardly obtain.  However, it is worth noting
that the functions $B_i$ may be expressed using determinants of parts
of the metric involving only $\ud \phi_i$, namely
\ben
B_3 = H^{-2/5} g_{\phi_3 \phi_3} , \quad B_2 = H^{-2/5} \frac{\det
  g_{(\phi_2, \phi_3)}}{g_{\phi_3 \phi_3}} , \quad B_1 = H^{-2/5}
\frac{\det g_{(\phi_1, \phi_2, \phi_3)}}{\det g_{(\phi_2, \phi_3)}} .
\een

There are CTCs if any $B_i$ is negative; the determinants appear in
  the expressions for $B_i$ as a manifestation of the standard result
  that a quadratic form is positive if and only if each of the leading minors is positive.  From the $g_{tt}$ coefficient, we have
\bea
- f^2 & = & - \frac{r^2 y^2 z^2 RYZ}{H^2 \prod_{i < j} (a_i^2 - a_j^2)^2
  B_1 B_2 B_3} + B_3 [g (1 + v_{32} + v_{31}) - v_{30}]^2 \nnr
&& + B_2 [g (1
+ v_{21}) - v_{20}]^2 + B_1 (g - v_{10})^2 .
\label{B123b}
\eea
Since $R = 0$ at the horizon and the left hand side is negative semi-definite, we generally
have some $B_i$ negative near the horizon, and so the solution
generally possesses naked CTCs.  There are, however, two special cases for which naked CTCs do not occur, which we now discuss.

\subsubsection{Supersymmetric black holes}

One way to obtain solutions free from naked CTCs is to demand that $f = 0$ at the horizon, which leads to the further condition
\ben
q = - \frac{\Xi_{1-} \Xi_{2-} \Xi_{3-} (a_1 + a_2) (a_2 + a_3) (a_3 + a_1) }{(1 - a_1 g - a_2 g - a_3 g)^2 g} .
\een
We then have the simplification
\ben
f = H^{-4/5} \left( 1 - \frac{(r_0^2 + y^2) (r_0^2 + z^2)}{(r^2 + y^2) (r^2 + z^2)} \right) ,
\een
where $r_0^2 = r_\text{h}^2$ is given by (\ref{r1}) and denotes the location
of a horizon.  At the horizon, we have $R = 0$, and at a horizon with
non-zero area, $B_1 B_2 B_3 \neq 0$.  Therefore from (\ref{B123b}), if
$f = 0$ at the horizon, then each of $g (1 + v_{32} + v_{31}) -
v_{30}$, $g (1 + v_{21}) - v_{20}$ and $g - v_{10}$ must vanish at the
horizon.  Differentiating (\ref{B123b}) with respect to $r$, we see
that $R'$ also vanishes at the horizon, so the radial function $R$
must possess a double root, and we find that it takes the form
\ben
R = \frac{(r^2 - r_0^2)^2}{r^2} \left( g^2 r^4 + [1 + (a_1^2 + a_2^2 +
  a_3^2) g^2 + 2 g^2 r_0^2] r^2 +
\frac{(a_1 a_2 a_3 - qg)^2}{r_0^4} \right) .
\een
There should be no other horizons outside $r = r_0$ to avoid naked CTCs, and this is guaranteed by positive $r_0^2$.

We then need to verify that each $B_i$ is non-negative outside the
horizon, which will place contraints on the parameters in terms of inequalities.  The expressions for $B_i$ are rather complicated, and so we
do not provide a full analysis.  However, if we choose each $a_i$ to be
positive but otherwise arbitrary, then one may verify that by taking
$g$ negative but with $|g|$ sufficiently small we obtain an example of
a solution free of naked CTCs.

The thermodynamical quantities simplify to
\bea
E & = & - \frac{\pi^2 \prod_{k < l} (a_k + a_l) [ \sum_i \Xi_i +
  \sum_{i < j} \Xi_i \Xi_j - (1 + a_1 a_2 a_3 g^3) (2 + \sum_i a_i g +
  \sum_{i < j} a_i a_j g^2) ]}{8 \Xi_{1+}^2 \Xi_{2+}^2 \Xi_{3+}^2 (1 -
  a_1 g - a_2 g - a_3 g)^2 g} , \nnr
S & = & - \frac{\pi^3 (a_1 + a_2) (a_2 + a_3) (a_3 + a_1) (a_1 a_2 +
  a_2 a_3 + a_3 a_1 - a_1 a_2 a_3 g)}{4 \Xi_{1+} \Xi_{2+} \Xi_{3+} (1 - a_1 g - a_2 g - a_3 g)^2 g r_0} , \nnr
J_i & = & - \frac{\pi^2 (a_1 + a_2) (a_2 + a_3) (a_3 + a_1) [a_i -
  (a_i^2 + 2 a_i \sum_{j \neq i} a_j + \prod_{j \neq i} a_j) g + a_1
  a_2 a_3 g^2] }{8
  \Xi_{1+} \Xi_{2+} \Xi_{3+} \Xi_{i+} (1 - a_1 g - a_2 g - a_3 g)^2 g}
, \nnr
Q & = & - \frac{\pi^2 (a_1 + a_2) (a_2 + a_3) (a_3 + a_1)}{2 \Xi_{1+}
  \Xi_{2+} \Xi_{3+} (1 - a_1 g - a_2 g - a_3 g) g} , \nnr
T & = & 0 , \quad \Omega_i = -g , \quad \Phi = -1 .
\eea

If $a = a_1 = - a_2 = - a_3$, then $r_0^2 = - a^2$ and so $f =
H^{-4/5}$.  Therefore $-f^2 = - H^{-8/5}$ is negative definite, and so
there cannot be supersymmetric black holes.  The absence of
supersymmetric black holes in this case, and analogously in five
dimensional supergravity with rotation parameters $a_1 = - a_2$, was
noted in the analysis of \cite{rotbhgaugedsugra}, there referred to as
type A.

We now consider whether the supersymmetric black holes can preserve
more than $\frac{1}{8}$ supersymmetry.  First, we consider the
vanishing of $\lambda_{1+}$, in which case (\ref{lambda1+}) holds.
Then $r_0^2 < 0$, so there are no such solutions.  There are two
possibilities (\ref{lambda1-a}, \ref{lambda1-b}) for the eigenvalue
$\lambda_{1-}$ vanishing, however the first, (\ref{lambda1-a}), is
trivial for supersymmetric black holes and satisfied only for
$\textrm{AdS}_7$, so we consider the second, (\ref{lambda1-b}).  Then
$(a_1 + a_2 + a_3) g < 2$, but then it is not possible to satisfy both
$(a_1 + a_2 + a_3) g < 0$ for the BPS constraint and $r_0^2 > 0$, so
again there are no such solutions.

\subsubsection{Topological solitons}

The second way of avoiding naked CTCs is to demand that $B_1 B_2 B_3$
also vanishes at the outermost root of $R$, so then the spacetime has
some minimum radius at which the geometry remains smooth, giving rise
to a topological soliton.  From (\ref{areadet}) and the expression for
the radial function $R$, we find that topological solitons occur if
the BPS constraint is supplemented by
\bea
q & = & \frac{\prod_i \Xi_{i-} [a_i - (a_i^2 + 2 a_i
  \sum_{j \neq i} a_j + \prod_{j \neq i} a_j ) g + a_1 a_2 a_3 g^2 ]
}{(1 - a_1 g - a_2 g - a_3 g)^4 g} \nnr
& = & \frac{\prod_i \Xi_{i-} (a_i - g r_{\textrm{h}}^2)}{(1 - a_1 g - a_2 g - a_3 g) g} .
\eea
The geometry ends at $r = r_0$ with
\ben
r_0^2 = - \frac{(a_1 a_2 + a_2 a_3 + a_3 a_1 - a_1 a_2 a_3 g)^2
  g^2}{(1 - a_1 g - a_2 g - a_3 g)^2} = - g^2 r_{\textrm{h}}^4 .
\label{r0susyts}
\een
Since there is no horizon, it is not necessary to demand that $r_0^2 >
0$, unlike the case of supersymmetric black holes, where it was needed
to ensure that the horizon area and entropy were real.  Since our
expression for $r_0^2$ is not positive, it is convenient to define a
new radial coordinate $\hat{r}^2 = r^2 - r_0^2$, which takes
values $0 \leq \hat{r} < \infty$.

In general, there is a conical singularity at $\hat{r} = 0$, as may be seen from a relevant part of
the metric,
\bea
\ud s^2 & = & H^{2/5} \left( \frac{(r^2 + y^2) (r^2 + z^2)}{R} \ud r^2 +
  B_1 (\ud \phi_1 + v_{10} \ud t)^2 \right) + \ldots \nnr
& = & H^{2/5} (r_0) (r_0^2 + y^2) (r_0^2 + z^2) \nnr
&& \bigg( \frac{(1 - a_1 g - a_2 g - a_3 g)^4 \ud
    \hat{r}^2}{\Xi_{1-} \Xi_{2-} \Xi_{3-} C (a_1 a_2 + a_2 a_3 + a_3 a_1 - a_1 a_2 a_3 g)^2} \nnr
&& ~ + \frac{C \Xi_{2-} \Xi_{3-} (a_2 + a_3 a_1 g)^2 (a_3 + a_1 a_2
  g)^2 \hat{r}^2 (\ud \phi_1 + v_{10} \ud t)^2}{\Xi_{1+}^2 \Xi_{1-} (1 - a_1 g - a_2 g - a_3 g)^6
  (r_0^2 + a_2^2)^2 (r_0^2 + a_3^2)^2 } 
\bigg) + \ldots ,
\eea
where
\bea
C & = & 1 - 5 \sum_i a_i g + 7 \sum_i a_i^2 g^2 + 19 \sum_{i < j} a_i a_j g^2 - 3 \sum_i a_i^3 g^3 - 19 \sum_{i \neq j} a_i^2 a_j g^3 \nnr
&& - 51 a_1 a_2 a_3 g^3 + 5 \sum_{i \neq j} a_i^3 a_j g^4 + 14 \sum_{i < j} a_i^2 a_j^2 g^4 + 39 a_1 a_2 a_3 \sum_i a_i g^4 \nnr
&& - 3 a_1 a_2 a_3 \sum_i a_i^2 g^5 - 14 a_1 a_2 a_3 \sum_{i < j} a_i a_j g^5 + 4 a_1^2 a_2^2 a_3^2 g^6 .
\eea
To ensure that there is no conical singularity at $\hat{r} = 0$, since
$\phi_1$ has period $2 \pi$, we need the quantization condition
\ben
\left( \frac{\Xi_{2-} \Xi_{3-} C (a_2 + a_3 a_1 g) (a_3 + a_1 a_2 g)
    (a_1 a_2 + a_2 a_3 + a_3 a_1 - a_1 a_2 a_3 g)}{\Xi_{1+} (1 - a_1 g
    - a_2 g - a_3 g)^5 (r_0^2 + a_2^2) (r_0^2 + a_3^2)} \right) ^2 = 1.
\label{qc}
\een
The supersymmetric topological solitons of
\cite{rotbhgaugedsugra} are given by $a = a_1 = a_2 = a_3$, there
known as type B, and $a = a_1 = - a_2 = - a_3$, there known as type
A.  For both cases, the quantization condition (\ref{qc}) cannot hold
for any rotation parameter with $-1 < ag < 1$ and with either $(a_1 +
a_2 + a_3) g < 0$ or $(a_1 + a_2 + a_3) g > 2$.  One could instead
consider making $\phi_1$ have period $2 \pi / k$ instead, for some
positive integer $k$, leading to solutions that are asymptotically
$\textrm{AdS}_7 / \mathbb{Z}_k$, in which case it becomes possible for
a quantization condition to hold with $-1 < a g < 1$.  We now consider whether the supersymmetric topological solitons can
preserve more than $\frac{1}{8}$ supersymmetry, for example the case
$a = a_1 = - a_2 = - a_3$ considered in \cite{rotbhgaugedsugra} was
shown to be $\frac{3}{8}$ supersymmetric.  We should again check, as
we previously did with the supersymmetric black holes, whether the
rotation parameters lie within the allowed
ranges to ensure the correct signature, so that the BPS constraint
is satisfied, and so that there are no naked singularities.
Additionally, we should check that we can find rotation parameters for
which the quantization condition (\ref{qc}) holds.  The eigenvalue
$\lambda_{1+}$ vanishes if (\ref{lambda1+}) holds.  It is possible to
find rotation parameters with $-1 < a_2 g < 1$ and $-1 < a_3 g < 1$ so
that $(a_1 + a_2 + a_3) g < 0$, but then it is not possible to satisfy
$-1 < a_1 g < 1$.  The other possibility of satisfying the BPS
constraint, of $(a_1 + a_2 + a_3) g > 2$, does not occur.  This
argument could have been used instead to rule out $\lambda_{1+} = 0$
for the supersymmetric black holes, since it does not rely on any
expression for $r_0$.  We next
consider the two possibilities (\ref{lambda1-a}, \ref{lambda1-b}) for
the eigenvalue $\lambda_{1-}$ vanishing.  If we have
(\ref{lambda1-a}), then we may choose $a_2 g$ and $a_3 g$ so that the
BPS constraint is satisfied through $(a_1 + a_2 + a_3) g < 0$.
However, we then find that $r_0^2 + \min(a_2^2, a_3^2) < 0$, and so a naked singularity cannot be avoided.  We
therefore move on to the second possibility that leads to vanishing
$\lambda_{1-}$, (\ref{lambda1-b}).  Setting $a_3 = - a_1$ with $a_2$
independent gives $\frac{1}{4}$ supersymmetric topological solitons;
we find that the rotation parameters may be chosen so that we have the
correct signature, a smooth geometry, satisfy the BPS condition, and
avoid naked CTCs.  The case $a = a_1 = - a_2 = - a_3$, which preserves
$\frac{3}{8}$ supersymmetry, as noted above, cannot be asymptotically
$\textrm{AdS}_7$, and satisfy both the BPS condition and the quantization
condition.

\section{Discussion}

We have obtained a black hole solution of seven dimensional gauged
supergravity with arbitrary angular momenta and equal $\textrm{U}(1)$
charges in the $\textrm{U}(1)^2$ truncation of the full
$\textrm{SO}(5)$ gauge group, complementing the solution of the
ungauged theory with arbitrary angular momenta and arbitrary charges
\cite{nearBPSsat}, and the solution of the gauged theory with equal
angular momenta and arbitrary charges \cite{crotbh7d}.  It remains to
find a general black hole solution of the gauged theory with arbitrary
angular momenta and arbitrary charges.

We have demonstrated similarities between some black hole solutions of
gauged supergravity theories in various dimensions in the case of
certain combinations of charges being set equal.  These may serve
as a guide to obtaining general black hole solutions of four and five
dimensional gauged supergravity with arbitrary angular momenta and
arbitrary charges as well.

\section*{Acknowledgements}

I would like to thank Malcolm Perry for helpful discussions.  This work has been supported by STFC.

\end{document}